\documentclass[twocolumn,amsmath,amssymb,prl]{revtex4}

\usepackage{epsfig}

\def\al{\alpha}
\def\be{\beta}

\def\de{\delta}

\def\th{\theta}

\def\ta{\tau}

\def\ph{\phi}

\def\ch{\chi}

\def\om{\omega}
\def\Ga{\Gamma}
\def\De{\Delta}

\def\Om{\Omega}

\def\fr#1#2{{{#1} \over {#2}}}
\def\half{{\textstyle{1\over 2}}}

\def\lsim{\mathrel{\rlap{\lower4pt\hbox{\hskip1pt$\sim$}}
    \raise1pt\hbox{$<$}}}
\def\gsim{\mathrel{\rlap{\lower4pt\hbox{\hskip1pt$\sim$}}
    \raise1pt\hbox{$>$}}}

\def\etal{{\it et al.}}

\newcommand{\beq}{\begin{equation}}
\newcommand{\eeq}{\end{equation}}
\newcommand{\bea}{\begin{eqnarray}}
\newcommand{\eea}{\end{eqnarray}}
\newcommand{\bit}{\begin{itemize}}
\newcommand{\eit}{\end{itemize}}
\newcommand{\rf}[1]{(\ref{#1})}
\def\Re{\hbox{Re}\,}
\def\Im{\hbox{Im}\,}
\newcommand{\nn}{\nonumber}

\def\mbf#1{\boldsymbol #1}

\def\nuTemplate#1#2#3#4{(#1^{(#2)})_{#3}^{#4}}

\def\aof#1#2{\nuTemplate{a_\text{of}}{#1}{#2}{}}
\def\aofs#1#2{\nuTemplate{a_\text{of}}{#1}{#2}{*}}
\def\ceff#1#2#3{\nuTemplate{c_\text{eff}}{#1}{#2}{#3}}
\def\ccombo#1{c^{(2)}_{#1}}

\def\N#1#2{{}_{#1}{\mathcal N}_{#2}}

\def\ari{{\ring{a}}}
\def\acfc#1#2{\ari^{(#1)}_{#2}}

\newcommand{\f}{f}
\newcommand{\geff}{g}

\def\ring#1{{\mathaccent'27 #1}}
\def\frac#1#2{\textstyle{{#1}\over{#2}}}

\def\thmax{\th_0}

\def\wt{\om_\oplus T_\oplus}
\def\omz{\De T}
\def\omc{\De T_c}

\begin{document}

\title{Relativity violations and beta decay}

\author{Jorge S. D\'iaz, V.\ Alan Kosteleck\'y, and Ralf Lehnert}

\affiliation{Physics Department, Indiana University,
Bloomington, IN 47405, U.S.A.}

\date{IUHET 573, May 2013; 
published as Phys.\ Rev.\ D {\bf 88}, 071902(R) (2013)}

\begin{abstract}

In some quantum theories of gravity,
deviations from the laws of relativity 
could be comparatively large while escaping detection to date.
In the neutrino sector,
precision experiments with beta decay yield new and improved constraints 
on these countershaded relativity violations.
Existing data are used to extract bounds of $3 \times 10^{-8}$ GeV 
on the magnitudes of two of the four possible coefficients,
and estimates are provided of future attainable sensitivities
in a variety of experiments. 

\end{abstract}

\maketitle

Investigating the behavior of neutrinos has yielded deep physical insights
since Pauli predicted their existence in 1930
to rescue the conservation of energy in beta decay
\cite{pauli}.
More recently,
numerous experiments have accumulated compelling evidence 
for neutrino oscillations, 
thereby confirming the existence of physics beyond the Standard Model (SM)
\cite{pdg}.
In this work,
we investigate the prospects for using beta decay
to search for another popular type of neutrino physics beyond the SM,
namely violations of Lorentz symmetry
in neutrino propagation. 
These violations 
could originate in a Planck-scale unification of quantum gravity
with other forces such as string theory
\cite{ksp},
and they are expected to be heavily suppressed
in effective quantum field theory 
describing physics at accessible energy scales,
typically by a factor involving the tiny ratio $m_W/m_P$
of the electroweak to Planck scales. 
 
The general framework describing deviations from Lorentz symmetry
in realistic effective quantum field theory 
is the Standard-Model Extension (SME)
\cite{sme}.
All quantum field operators for Lorentz violation involved 
in the propagation of neutrinos have been classified and enumerated
\cite{km12}.
Most of these can be studied using neutrino oscillations,
which compare the way different neutrinos propagate
and provide interferometric sensitivity to energy differences
between neutrinos 
\cite{LVexpt}.
Some effects cannot be detected by neutrino oscillations 
because they are produced by `oscillation-free' operators 
that change all neutrino energies equally.
Most of these can instead be studied 
by comparing neutrino propagation to other species,
such as time-of-flight experiments matching 
the group velocity of neutrinos with that of photons.
However,
four oscillation-free operators leave unaffected 
the neutrino group velocity and so cannot be detected in this way.
Instead,
they must be accessed through 
physical processes that involve 
neutrino phase-space properties,
such as quantum decays.
These operators are rare examples of `countershaded' Lorentz violations
\cite{kt}:
relativity-violating effects that could be enormous 
compared to ones suppressed by the ratio $m_W/m_P$
and that nonetheless could have escaped detection to date. 
These could provide an interesting path for building models
with viable Lorentz violation
obviating the typical requirement of a heavy suppression factor.
The present work focuses on methods to constrain 
these unique effects.

The four countershaded neutrino operators 
are of renormalizable mass dimension $d=3$,
are odd under CPT,
and are controlled by coefficients conventionally denoted
$\aof 3 {jm}$,
where $j,m$ are angular-momentum quantum numbers with $j=0,1$.
Conservation of energy and momentum is assured
by taking these four coefficients to be constant
as usual for couplings beyond the SM, 
so all physics other than Lorentz and CPT violation is conventional.
Dimensional arguments suggest these coefficients 
are likely to dominate at accessible energies
and can be measured sensitively in low-energy processes.
Here,
we demonstrate that experiments on beta decay,
a well-studied low-energy process
with the potential for precision measurement,
provide high sensitivity 
to these oscillation-free effects.
We use available data to improve the current limit
on $\aof 3 {00}$ by over an order of magnitude 
and to obtain a first measurement of $\aof 3 {10}$.
We show that targeted existing and forthcoming experiments 
can access all four coefficients
and further improve sensitivities,
thereby revealing an effect
or substantially reducing the window for countershaded Lorentz violation.

For a beta decay involving an antineutrino of mass $m_\nu$ 
and 4-momentum $q^\al=(\om,{\mbf q})$,
the antineutrino phase space can be written as $d^3q=f(\om)d\om\,d\Om$, 
where the antineutrino function 
$f(\om) \approx \om^2 - \half m_\nu^2 -2\om \de \om $ 
encodes the Lorentz-violating modifications
\beq
\de \om = 
-\sum_{jm} e^{im\om_\oplus T_\oplus} \N{}{jm} \aof{3}{jm}
\label{fw}
\eeq 
arising from the modified antineutrino dispersion relation
\cite{km12}
$\om = |\mbf q| + m_\nu^2/|\mbf q| + \de \om$.
In Eq.\ \rf{fw}, 
the sidereal time $T_\oplus$ controls 
the harmonic variation of the antineutrino function in the laboratory 
produced by the Earth's sidereal rotation at frequency 
$\om_\oplus \simeq 2\pi$/(23 h 56 min).
The factors $\mathcal{N}_{jm}$ 
contain information about the direction of propagation 
of the antineutrinos,
expressed relative to the canonical Sun-centered frame 
\cite{sunframe}. 
Denoting the electron mass by $m_e$ 
and its 4-momentum by $p^\al=(E,{\mbf p})$,
the differential spectrum for a single beta decay is given by 
$d\Ga/dT = C(T)\int d\Om \f(T_0-T)$,
where $C(T)$ is a function
of the electron kinetic energy $T=E-m_e$
and $T_0$ is the conventional endpoint energy for $m_\nu=0$.
For simplicity,
we assume measurable Lorentz violation is limited to the neutrino sector.
This choice is compatible with existing constraints on other species,
including restrictions from electroweak symmetry
\cite{km04},
and with observability requirements imposed by standard field redefinitions
\cite{sme}.
Early theoretical works considering 
neutrino Lorentz violation in beta decay 
include Refs.\  \cite{bbgescc}.
Lorentz-violating effects arising from weak interactions in beta decay 
without neutrino Lorentz violation are studied in Ref.\ \cite{nwt},
while CPT violation without Lorentz violation in double beta decay is
considered in Ref.\ \cite{bbbk}.

First, consider precision experiments designed to detect neutrino mass 
directly by studying the endpoint of tritium beta decay.
Recent experimental measurements using tritium have been performed 
in Troitsk \cite{troitsk,troitsk2,troitsk3} and 
Mainz \cite{mainz},
while the next-generation 
Karlsruhe Tritium Neutrino experiment (KATRIN) \cite{katrin} 
is expected to begin taking data shortly.
In these experiments,
beta-decay electrons are guided by a magnetic field 
from the decay region to an electrostatic filter, 
where electrons with energies 
near the endpoint $T_0\simeq 18.6$ keV are selected. 
The electrons guided to the filter
emerge from the decay region within a solid angle $\De\Om$
measured about the $z$ axis along the magnetic field,
which we take as horizontal,
with the acceptance cone having angular aperture $\thmax$.
For this configuration,
the orientation factors $\N{}{jm}$ in Eq.\ \rf{fw} become 
\beq
\N{}{jm} = 
\sum_{m'm''} Y_{jm'}(\th,\ph)  
d^{(j)}_{m'm''}(-\pi/2) e^{-im''\xi} d^{(j)}_{m''m}(-\ch),
\label{njm}
\eeq
where $Y_{jm'}(\th,\ph)$ are spherical harmonics in the laboratory frame, 
$d^{(j)}_{m'm''}$ are the little Wigner matrices,
$\xi$ is the angle of the magnetic field at the source 
measured east of local north, 
and $\ch$ is the colatitude of the laboratory.

The selection of electrons lying within $\De\Om$ 
introduces experimental sensitivity to direction-dependent effects
arising from neutrino Lorentz violation
and hence provides access in principle to all four coefficients
$\aof 3 {jm}$. 
For $m=\pm 1$, 
the phase factor in Eq.\ \rf{fw} varies sinusoidally 
in the sidereal time $T_\oplus$,
so time stamps for the data permit a search for sidereal variations
and hence measurements of 
$\aof 3 {11}$ and $\aof 3 {1-1}\equiv -\aofs 3 {11}$ or,
equivalently,
of $\Re \aof 3 {11}$and $\Im \aof 3 {11}$. 
For $m=0$ there is no time dependence,
but the Lorentz violation modifies the shape 
of the differential beta spectrum 
and so a study of the time-averaged spectral shape 
instead can enable measurements of $\aof 3 {00}$ and $\aof 3 {10}$.

In this work,
we use published results from the Troitsk and Mainz experiments 
to place conservative constraints on the coefficients $\aof 3 {j0}$
and others,
and we estimate sensitivities 
attainable in principle 
from the unpublished raw data in these experiments and 
in KATRIN.
In practice,
data for the tritium endpoint spectrum are available 
only over a small energy range $\omc = T_0-T_c$ 
from some cutoff energy $T_c$ to $T_0$,
so the Lorentz-violating modifications to the spectral shape
are only partly observable.
For this energy range,
the decay rate takes the form
\beq
\fr{d\Ga}{dT} \approx 
B + C [(\omz + k(T_\oplus))^2 - \frac 12 m_\nu^2],
\label{rate}
\eeq
where 
$B$ is the experimental background rate,
$C$ is approximately constant,
and $\omz = T_0 - T$.
The function $k(T_\oplus)$ 
contains the SME coefficients,
\bea
k(T_\oplus) &=& 
\frac 1 {\sqrt{4\pi}}
\aof 3 {00}
- \sqrt{\frac {3} {4\pi}}
\cos^2 \hskip-2pt \half\thmax 
\sin \ch \cos\xi ~\aof 3 {10}
\nn\\
&&
- \sqrt{\frac {3} {2\pi}}
\cos^2 \hskip-2pt \half\thmax 
[ \sin\xi ~\Im (\aof 3 {11} e^{i\wt})
\nn\\
&&
\hskip 25pt
+ \cos\ch \cos\xi ~\Re (\aofs 3 {11} e^{-i\wt}) ].
\eea
The result \rf{rate} reveals that $k(T_\oplus)$ 
acts to shift the tritium endpoint spectrum along the energy axis
without changing its shape,
independent of the value of $m_\nu^2$. 
Sensitivity to this effect therefore requires experimental access
to absolute energy measurements.
The shift can be positive or negative
and depends in part on the location and orientation of the experiment
and on the sidereal time.
The coefficients $\Re \aof 3 {11}$ and $\Im \aof 3 {11}$
induce harmonic oscillations of the spectrum 
along the energy axis at frequency $\om_\oplus$,
while $\aof 3 {00}$ and $\aof 3 {10}$
shift the location of the endpoint relative to the usual case.
For data collected over a long period
the harmonic oscillations average away,
and so only the coefficients 
$\aof 3 {00}$ and $\aof 3 {10}$
produce observable effects.
For simplicity in what follows,
we take only one nonzero coefficient at a time,
noting that fortuitous cancellations cannot simultaneously occur 
in experiments with different values of $\ch$, $\xi$, $\th_0$. 

Conservative constraints on the coefficients $\aof 3 {j0}$
can be placed using published results and 
the time-averaged form of Eq.\ \rf{rate}.
Consider first the Troitsk experiment 
\cite{troitsk,troitsk2,troitsk3}.
The experiment is located at a colatitude $\chi\simeq35^\circ$,
has decay-pipe axis pointing barely west of north with
$\xi\simeq-5^\circ$,
and has acceptance-cone aperture 
$\thmax\simeq20^\circ$
\cite{vsp}. 
The averaged endpoint energy measured in this experiment
is 18576 eV,
which cannot be taken as an absolute energy measurement
\cite{troitsk3}
but lies within about 2 eV of the expected endpoint.
Taking $\pm 5$ eV as the upper limit of a possible constant shift
yields the constraints
$|\aof 3 {j0}| < 2\times10^{-8}$ GeV
for both $j=0$ and $j=1$.

The Mainz experiment
reports a series of 12 measurements of the endpoint energy 
under different experimental conditions,
denoted Q1-Q12
\cite{mainz}.
The apparatus was located at colatitude $\chi\simeq40^\circ$
and the axis of the decay pipe had 
orientation $\xi\simeq-65^\circ$ relative to local north 
\cite{cw}.
The theoretical maximum value for the electron kinetic energy
for this experiment is $T_0 = 18574.3\pm 1.7$ eV
\cite{mainz}.
For definiteness consider run Q12,
which involves a wider acceptance-cone aperture 
$\thmax\simeq 62^\circ$
and has measured endpoint $18576.6\pm 0.2$ eV.
Interpreting the difference between the measured and theoretical endpoints
via Eq.\ \rf{rate} gives
$\aof 3 {00} = 8.2\pm6.1\times10^{-9}$ GeV
and $\aof 3 {10} = -2.4\pm1.8\times10^{-8}$ GeV.
Constraints at similar levels can be obtained
from the other runs.
More conservatively,
we can infer that a constant shift of $\pm 5$ eV would be observable,
which for this experiment leads to the constraints 
$|\aof 3 {00}| < 2\times10^{-8}$ GeV
and $|\aof 3 {10}| < 5\times10^{-8}$ GeV.

Taken together,
the above results permit us safely to conclude that 
$|\aof 3 {j0}| \lsim 3 \times 10^{-8}$ GeV 
for both $j=0$ and $j=1$
at better than a 90\% confidence level. 
Despite their conservative nature,
these constraints significantly improve 
on existing limits
\cite{tables},
which have been extracted from studies of threshold effects.
Threshold effects have been investigated 
only in the purely isotropic model,
for which $\acfc 3{} \equiv \aof 3 {00}/\sqrt{4\pi}$ 
is the sole nonzero coefficient for $d=3$.
The best existing constraint 
$|\acfc 3{}|<1.9\times 10^{-7}$ GeV
is obtained using IceCube data 
\cite{km12},
so the result for $\aof 3 {00}$
presented here represents an improvement
of more than an order of magnitude.
Our constraint on $\aof 3 {10}$ is the first in the literature.

Dedicated analyses of the raw data by the Troitsk, Mainz, 
and KATRIN collaborations
could improve on these constraints.
For the Troitsk and Mainz experiments,
sensitivities of 
$|\aof 3 {j0}| \lsim 10^{-9}$ GeV
or better appear attainable
by focusing attention on absolute energies.
For KATRIN 
\cite{katrin},
the apparatus is located at
colatitude $\chi\simeq41^\circ$, 
has orientation relative to local north of $\xi\simeq16^\circ$,
and has acceptance-cone aperture $\thmax\simeq 51^\circ$.
With 30 days of data,
statistical confidence levels 
suggest a reach about two orders of magnitude
beyond the new constraints reported above.
In all these experiments,
first measurements 
of $\Re \aof 3 {11}$ and $\Im \aof 3 {11}$ 
could be achieved by binning in sidereal time
and fitting to Eq.\ \rf{rate}.

We remark in passing that tritium endpoint experiments
also have sensitivity to the full spectrum 
of neutrino SME coefficients.
Competitive constraints arise 
for effects that dominate at low energies
and hence for operators of low $d$.
For example,
the $d=2$ coefficients $\ceff 2 {1m}{ab}$,
where $a,b = e,\mu,\ta$ are flavor indices and $m=0,\pm 1$,
control helicity-flipping CPT-even Lorentz violation
\cite{km12}.
Calculation reveals that linear combinations
$\ccombo m$ of these coefficients
act to shift the squared mass in tritium beta decay,
so the endpoint spectrum is governed by an effective squared mass
$m_{\rm eff}^2 = m_\nu^2 + k_m \ccombo m$,
where $k_m$ depends on 
$\ch$, $\xi$, $\thmax$,
and also on $T_\oplus$ for $m=\pm1$
\cite{fn1}.
For instance,
for the Troitsk experiment 
$k_0 \simeq 0.5$,
and the reported measurement 
$m_{\rm eff}^2 = -0.67 \pm 2.53$ eV$^2$
translates into the bound 
$\ccombo 0 < 4\times10^{-18}$ GeV$^2$,
independent of $m_\nu^2$.
Using the standard mixing parameters
\cite{pdg}
$\theta_{12}=0.59$, 
$\theta_{13}=0.16$, 
$\theta_{23}=0.79$, 
assuming zero conventional CP phase $\de$,
and taking only one coefficient $\ceff 2 {10}{ab}$
at a time,
this bound translates into six constraints 
in units of $10^{-17}$ GeV$^2$:
$-1 < \ceff 2{10}{ee}$,
$-2 < \ceff 2{10}{\mu\mu}$,
$-2 < \ceff 2{10}{\ta\ta}$,
$\Re\ceff 2{10}{e\mu}< 1$,
$-3 < \Re\ceff 2{10}{e\ta}$
and 
$\Re\ceff 2{10}{\mu\ta}< 1$.
Similarly,
for the Mainz experiment 
$k_0 \simeq 0.2$,
and the reported combined limit 
$m_\nu^2=-0.6 \pm 3.0$ eV$^2$
yields the constraint
$\ccombo 0 < 1\times10^{-17}$ GeV$^2$,
which gives the six constraints
$-2 < \ceff 2{10}{ee}$,
$-4 < \ceff 2{10}{\mu\mu}$,
$-5 < \ceff 2{10}{\ta\ta}$,
$\Re\ceff 2{10}{e\mu} < 3$,
$-8 < \Re\ceff 2{10}{e\ta}$
and 
$\Re\ceff 2{10}{\mu\ta}< 3$,
all in units of $10^{-17}$ GeV$^2$.
These are first results for $\ceff 2 {10}{ab}$
for all flavors $ab$ except $e\mu$.
For KATRIN $k_0 \simeq 0.5$,
and a sensitivity of $0.04$ eV$^2$
corresponds to a competitive reach of
$\ccombo 0 < 8\times10^{-20}$ GeV$^2$,
offering an improvement of about two orders of magnitude.
Sidereal analyses for these experiments could
yield $m_\nu^2$-independent two-sided constraints 
on the coefficients $\ceff 2 {1\pm1}{ab}$.

Returning to countershading studies,
neutron decay \cite{ns} 
offers another interesting experimental option.
Precision experiments investigating the beta spectrum from neutron decay 
typically have lesser sensitivity
but could detect the full spectral distortion
instead of only the endpoint shift.
This includes energy regions 
where the coefficients $\aof 3 {jm}$ induce 
maximal deviation from the conventional spectrum,
which can lie comparatively far from the endpoint $T_0\simeq 0.78$ MeV
\cite{fn2}.

Consider,
for example,
experiments that measure only $T$.
The differential beta spectrum $d\Ga/dT$
must then be constructed
by integrating over the lepton directions of travel,
so only the isotropic coefficient 
$\acfc 3{}$ 
can play a role.
We find
$d\Ga/dT \propto F(Z,T) |\mbf p| (T+m_e) (\omz + \acfc 3{})^2$,
where $F(Z,T)$ is the Fermi function.
The conventional spectrum peaks at $T\simeq 0.25$ MeV,
while the residual Lorentz-violating spectrum 
has a maximum at $T_m \simeq 0.41$ MeV.
The ratio $R$ of the residual to the conventional spectra
at $T_m$ is $R \simeq 5\times 10^3 \acfc 3{}/$GeV.
An experiment with a plausible sensitivity of 0.1\% 
in this energy region would thus have an estimated reach of 
$|\acfc 3{}| < 2\times 10^{-7}$ GeV,
comparable to the constraint from IceCube 
\cite{km12}.

Some neutron-decay experiments are designed to measure the
antineutrino-electron correlation parameter $a$
associated with the angle between 
the phase velocities of the two emitted leptons.
Denoting by $N_+$ the number of parallel leptons at energy $T$
and by $N_-$ the number of antiparallel ones, 
a standard observable is the asymmetry 
$a_{\rm exp} = (N_+ - N_-)/(N_+ + N_-)$,
which in the absence of Lorentz violation is $a_{\rm exp} = a|\mbf p|/E$. 
The correction involves all $j=1$ coefficients $\aof 3 {jm}$,
including a signal from sidereal variations 
associated with $\aof 3 {11}$. 
Assuming a plausibly attainable 0.1\% measurement 
of $a_{\rm exp}$ near 0.35 MeV
for an experiment located at $\ch=45^\circ$
and taking $a = -0.103$ 
\cite{pdg},
we find estimated reaches of
$|\aof 3{10}| < 5\times 10^{-8}$ GeV
and 
$|\Re \aof 3{11}|, |\Im \aof 3{11}| < 4\times 10^{-8}$ GeV.

Another class of experiments seeks to measure
the correlation parameter $B$ controlling the 
angle between the neutron spin and the antineutrino phase velocity.
This requires sensitivity to the neutron polarization
and reconstruction of the antineutrino emission.
A standard observable is 
$B_{\rm exp} = (N_{--} - N_{++})/(N_{--} + N_{++})$,
where $N_{--}$ and $N_{++}$ count events 
with both the electron and the proton emitted 
against and along the direction of the neutron spin,
respectively.
The Lorentz-violating correction to the asymmetry 
involves all $j=1$ coefficients $\aof 3 {jm}$,
along with a dependence on $a$, $B$,
and the spin-electron correlation parameter $A$.
For a plausible 0.1\% measurement of $B_{\rm exp}$ near 0.35 MeV
at $\ch=45^\circ$
and taking $a = -0.103$, $B = 0.9807$, $A = -0.1176$ 
\cite{pdg},
we obtain estimated attainable sensitivities 
$|\aof 3{10}| < 2\times 10^{-6}$ GeV
and $|\Re \aof 3{11}|, |\Im \aof 3{11}| < 1\times 10^{-6}$ GeV.

Single beta decays of other nuclei and particles such as pions
\cite{ba},
kaons, or muons 
are also worth studying and are likely to yield similar limits
on $\aof 3 {jm}$.
A qualitatively different option is provided by double beta decay.
This is a second-order weak process,
so a lesser sensitivity is generically to be expected.
However,
precision experiments searching for the neutrinoless mode
typically also generate a large sample of two-antineutrino events.
The corresponding statistical reach can be significant,
and all four $\aof 3 {jm}$ appear in angular correlations
if the electron directions can be reconstructed.

For the two-antineutrino mode,
we limit attention here to an analysis using only
the differential spectrum $d\Ga/dK$ of the summed energies $K = T_1 + T_2$ 
of the two emitted electrons.
This involves integration over the neutrino directions,
which implies only isotropic effects are observable
and hence that the residual spectrum depends 
only on $\acfc 3{}$.
Defining $\De K = K_0 -K $
where $K_0$ is the maximum kinetic energy available in the decay,
the sum electron spectrum $d\Ga/dK$ 
is modified by the factor $\De K^5 \to (\De K + 2\acfc 3 {})^5$,
revealing a distortion of the spectral shape.
Consider $^{136}$Xe,
for example,
for which $K_0 \simeq 2.46$ MeV and the maximum 
of the conventional sum electron spectrum  
occurs at energy $K \simeq 0.86$ MeV.
The residual spectrum has maximum at $K_m \simeq 1.02$ MeV,
so the ratio of the residual and conventional spectra at $K_m$ is 
$R \simeq 7\times 10^3 \acfc 3{}/$GeV.
An experiment with a plausible precision of 0.1\% near $K_m$
would hence have an estimated reach  of
$|\acfc 3{}| < 2\times 10^{-7}$ GeV.

In the neutrinoless mode of double beta decay, 
the neutrino is virtual and must have Majorana couplings 
to generate a nonzero amplitude.
The coefficients $\aof 3 {jm}$ are Dirac couplings,
so their contribution to the amplitude
must be suppressed by some other Majorana coupling
and competitive sensitivities are unlikely. 
However, 
numerous Lorentz-violating Majorana operators exist 
\cite{km12}.
Here,
we outline a few implications.
The usual half life is 
$T_{1/2}=(G|M|^2 m_\nu^2)^{-1}$, 
where 
$G$ is a known function of the nuclear radius $R$ and other quantities,
and $M$ is the nuclear matrix element.
Calculation shows leading-order effects can arise only from
CPT-odd Majorana operators controlled by the SME coefficients  
$\widehat g{}_{M+}^{\al\be}$
\cite{km12}.
Suppressing the orientation dependence 
and denoting the dimensionless effective coefficient by $\geff$,
we find the half-life corrections include the replacement
$m_\nu^2\to m_\nu^2+m_\nu\geff/R+(\geff/R)^2$.
This reveals the striking feature 
that Lorentz-violating neutrinoless double beta decay can occur 
even for negligible $m_\nu$
\cite{fgp},
with the role of the Majorana mass played by $\geff$.
Experiments placing an upper bound on $m_\nu^2$ 
can therefore also report a constraint on $\geff$.
For example,
the current limit on the neutrinoless double beta decay of $^{136}$Xe 
\cite{exo} 
corresponds to the constraint $|\geff|\lsim 10^{-9}$. 
Note that
experiments with different locations and orientations 
generically have different sensitivities 
due to the directional and sidereal dependences, 
as do experiments with different isotopes
due to the $R$ dependence
\cite{fn3}.

The results reported here 
demonstrate that studies of beta decay 
can achieve interesting sensitivities 
to countershaded neutrino Lorentz violation.
For gravitational Lorentz violation,
countershading has recently been eliminated to the keV scale
\cite{tables}.
Here,
the constraints obtained lie at the eV scale,
improving by more than an order of magnitude
the existing bound on one effect 
and setting a first bound on another.
Models involving these neutrino operators 
at scales comparable to $m_e$ are now excluded,
while the new constraints
must be satisfied even by models with effects 
at neutrino-oscillation scales.

This work was supported in part
by the Department of Energy
under grant DE-FG02-91ER40661
and by the Indiana University Center for Spacetime Symmetries.

\end{document}